\documentclass[conference]{IEEEtran}

\usepackage{setspace} 
\usepackage{amsmath}
\usepackage{amssymb}

\usepackage{array}
\usepackage{amsfonts}
\usepackage{amsthm} 
\newtheorem{theorem}{Theorem}

\newtheorem{remark}{Remark}

\usepackage{cite}
\usepackage{mathtools}
\usepackage{cases}
\usepackage{graphicx}
\usepackage{subfigure}
\usepackage [paperwidth=8.5in, paperheight=11in, width=7.0in, height=9in]{geometry}
\usepackage{xcolor}

\title{ Over-the-Air Computation Systems: Optimal Design with Sum-Power Constraint}
\author{Xin Zang, Wanchun Liu$^\dagger$, \emph{Member, IEEE}, Yonghui Li, \emph{Fellow, IEEE}, Branka Vucetic, \emph{Fellow, IEEE}}
\begin{document}
	\maketitle
\begin{abstract}
\let\thefootnote\relax\footnote{The authors are with School of Electrical and Information Engineering, The University of Sydney, Australia.
	Emails:	\{xin.zang,\ wanchun.liu,\ yonghui.li,\ branka.vucetic\}@sydney.edu.au. (\emph{Wanchun Liu is the corresponding author.})
	The works of W. Liu and B. Vucetic were supported by the Australian Research Council's Australian Laureate Fellowships Scheme under Project FL160100032. The work of Y. Li was supported by ARC under Grant DP190101988.
}Over-the-air computation (AirComp), 
which leverages the superposition property of
wireless multiple-access channel (MAC) and the mathematical tool of function representation, has been considered as a promising technique for effective collection and computation of massive sensor data in wireless Big Data applications.
In most of the existing work on AirComp, optimal system-parameter design is commonly considered under the peak-power constraint of each sensor.
In this paper, we propose an optimal transmitter-receiver (Tx-Rx) parameter design problem to minimize the computation mean-squared error (MSE) of an AirComp system under the sum-power constraint of the sensors. We solve the non-convex problem and obtain a closed-form solution. Also, we investigate another problem that minimizes the sum power of the sensors under the constraint of computation MSE.
Our results show that in both of the problems, the sensors with poor and good channel conditions should use less power than the ones with moderate channel conditions.
\end{abstract}


\section{Introduction}
For the implementation of Internet of Things (IoT)-based Big Data applications, there are two important challenges:
one is to wirelessly collect data from massive number of smart devices with restricted radio-frequency (RF) spectrum bandwidth, especially when the data requires real-time processing~\cite{IoTCritical,LiuJIoT,KangTWC,KangJIoT,KangICC,KangGC,LiuGC};
the other is the effective information fusion of massive data, i.e., an effective computation problem~\cite{wu2013data} and{\cite{NanZhao}}.

Over-the-air computation (AirComp), which leverages the superposition property of
wireless multiple-access channel (MAC) and the mathematical tool of function representation, is a promising technique to tackle the above challenges~\cite{GoldenbaumTCOM,GoldenbaumTSP,GoldenbaumWIOPT,Katabi,GuangxuMIMO,Wen19,liu2019over,Goldenbaumletter}.
Specifically, an AirComp system consists of $K$ sensors and one receiver, and the receiver aims to compute a pre-determined function of the sensors' measurement signals.
Each sensor of the AirComp system sends its pre-processed original signal simultaneously to the receiver through a MAC. Then, by applying a post-processing function on the received superimposed signal, the receiver directly obtains an estimation of the desired function output of the $K$ sensors' signals (see e.g.~\cite{liu2019over} for details).

The pre- and post-processing function design problems of AirComp systems have been comprehensively investigated in \cite{GoldenbaumTCOM,GoldenbaumTSP,GoldenbaumWIOPT,Katabi}. 
Most of the recent researches focus on optimal estimation of the sum of the $K$ pre-processed signals through a non-perfect MAC with unequal channel coefficients and non-zero
receiver noise~\cite{GuangxuMIMO,Wen19,liu2019over,Goldenbaumletter}.
In~\cite{GuangxuMIMO} and~\cite{Wen19}, the transmitting and receiving beamforming problems of multi-antenna AirComp systems were considered to minimize the estimation distortions.
In \cite{liu2019over}, 
the optimal single-antenna AirComp design and the scaling law analysis in terms of the number of sensors was investigated.
In \cite{Goldenbaumletter}, the estimation distortion of the sum signal under imperfect channel state information was analyzed.
More recently, AirComp has been applied to emerging mobile applications such as over-the-air consensus~\cite{consensus}, wireless cooperative computing~\cite{consensus2},  and wireless distributed machine learning~\cite{GuangxuWangyong,Gunduz,Osvaldo}.

In most of the existing work on AirComp systems, 
the optimal design is commonly considered under the
peak-power constraint of each sensor~(see e.g. \cite{GuangxuMIMO} and \cite{liu2019over}).
We note that the sum-power constrained AirComp system is also worth investigating for three reasons.
First,  the sum-power constrained conventional MAC systems have been studied extensively in the literature~\cite{Gupta,Boche,Wilson}.
Second, to enhance battery lives, the wireless sensors of an AirComp system can be wirelessly powered by a power beacon with a power constraint. Thus, the power beacon can decide how to distribute the power to the sensors, i.e., the sum power of the sensors for transmission is limited.
Last, the sum power of the sensors should be limited to meet the requirement of interference caused at the nearby in-band communication systems.

In this paper, we consider the optimal design of an AirComp systems under the sum-power constraint. The contributions are summarized as:
1) We formulate the optimal transmitter-receiver (Tx-Rx) scaling factor design problem to minimize the mean-squared error (MSE) of the estimation of the sum of the $K$ sensors' pre-processed signals under the sum-power constraint. We convert the original non-convex problem into a convex one and obtain the closed-form solution.
2) We consider another problem that minimizes the sum power of the AirComp system under the constraint of the AirComp MSE. A closed-form solution of the problem is also obtained.
3) Our results show some important properties of the optimal MSE and the optimal sum-power policies. For example, in both policies, the sensors with poor and good channel conditions should use less power for transmission than the ones with moderate channel conditions.

\section{System model}
We consider an AirComp system with $K$ sensors and a receiver, where each device is equipped with a single antenna.
Sensor $k$'s pre-processed signal is $x_{k}\in \mathbb{R},\forall k\in \mathcal{K}$, where $\mathcal{K}\triangleq\{1,\cdots,K\}$.  We assume that $x_k$ has normalized variance~\cite{GuangxuMIMO,liu2019over}.
Each sensor linearly scales its signal by a Tx-scaling factor, $b_k$, and sends $b_k x_k$ to the receiver simultaneously
via a multiple-access channel (MAC).
The channel coefficient between sensor $k$ and the receiver is $h_{k}$. 
The receiver linearly scales the received signal by the Rx-scaling factor $g$ as the computing output of sum of the original signals $\sum_{k=1}^{K} x_k$, and is given as~\cite{liu2019over}
\begin{equation}\label{signal}
{y}= g\left(\sum_{k=1}^{K} h_k b_k x_k +n\right),
\end{equation}
where $n$ is the receiver-side additive white Gaussian noise (AWGN) with zero mean and variance $\sigma^2$.
{Note that the Rx-scaling factor is designed for providing power compensation for the computation of $\sum_{k=1}^{K} x_k$ rather than for improving the signal-to-noise ratio (SNR).}

The computation distortion is measured by the estimation MSE of $\sum_{k=1}^{K} x_k$, and is given as
\begin{equation}\label{mse_1}
\mathsf{MSE} \triangleq \mathsf{E}\left[\vert y -\sum_{k=1}^{K} x_{k} \vert^2\right],
\end{equation}
{where $\mathsf{E}[\cdot]$ is the expectation operator.}
Substituting \eqref{signal} into \eqref{mse_1}, we have
\begin{equation}\label{mse}
\mathsf{MSE} = \sum_{k=1}^{K} \vert g h_k b_k -1 \vert^2 + \sigma^2 \vert g \vert^2.
\end{equation}
The sum power of the AirComp system is 
\begin{equation}
\mathsf{PW} \triangleq \sum_{k=1}^{K}\vert b_k \vert^2.
\end{equation}

We investigate the MSE minimization problem under the sum-power constraint and the sum-power minimization problem under the MSE constraint in terms of the Tx and Rx scaling factors, in the sequel.

\section{Optimal Computation MSE with Sum-Power Constraint}\label{sec:P1}
In this section, we consider the optimal scaling-factor design problem to minimize the computation MSE under the constraint of sum power. 
The problem is formulated as
\begin{subequations}\label{pro6}
	\begin{alignat}{2}
	&\!\min_{g,\{b_k\}}      &\qquad& \mathsf{MSE} = \sum_{k=1}^{K} \vert g h_k b_k -1 \vert^2 + \sigma^2 \vert g \vert^2\label{5a}\\
	\label{5b}
	&\text{subject to} &      & \mathsf{PW} = \sum_{k=1}^{K}\vert b_k \vert^2 \leq P,
	\end{alignat}
\end{subequations}
where $P$ is the sum-power constraint of the AirComp system, and $\{b_k\}$ denotes the set of $\{b_1,\cdots,b_K\}$.

Similar to that of the peak-power constrained problem in \cite{liu2019over}, given  the target function \eqref{5a}, and the complex
Rx-scaling factor $g$ and the channel coefficient $h_k$, one can adjust the phase of $b_k$  such that $g h_k b_k$ is real and non-negative and hence minimizes $|g h_k b_k -1|$ in \eqref{5a}. Thus, only the magnitudes of $g$,
$\{h_k\}$ and $\{b_k\}$ have effect on achieving the minimum MSE in problem \eqref{pro6}.
Without loss of generality, we assume that $g,h_k,b_k\in \mathbb{R}$, $h_k>0,\forall k\in\mathcal{K}$, in the rest of the paper.

{It is clear that \eqref{5b} is an active constraint since a larger $\vert b_k \vert$ leads to a smaller $\mathsf{MSE}$.	
However, problem (5) is non-convex due to the non-convexity of the target function \eqref{5a}. 
By letting $\hat{b}_k\triangleq g b_k,\forall k \in \mathcal{K}$, problem (5) can be converted to an equivalent problem as
\begin{subequations}\label{pro6'}
	\begin{alignat}{2}
	&\!\min_{g,\{\hat{b}_k\}}      &\qquad&  \sum_{k=1}^{K} \vert h_k \hat{b}_k -1 \vert^2 + \sigma^2 \vert g \vert^2\label{5a'}\\
	\label{5b'}
	&\text{subject to} &      &  \sum_{k=1}^{K}\vert \hat{b}_k \vert^2 = \vert g \vert^2 P.
	\end{alignat}
\end{subequations}
Taking \eqref{5b'} into \eqref{5a'}, the problem is converted to a convex problem as
\begin{equation}\label{prob7}
	\min_{\{\hat{b}_k\}}      \qquad \sum_{k=1}^{K} \vert h_k \hat{b}_k -1 \vert^2 + \frac{\sigma^2}{P} \sum_{k=1}^{K} \vert \hat{b}_k \vert^2.
\end{equation}
The solution of problem \eqref{prob7} is obtained straightforwardly by finding the extreme point of the target function. Then, we can have the following result.}

\begin{theorem}\label{theorem:1}
	\normalfont
The optimal Rx-scaling factor $g^\star$ and the optimal  Tx-scaling factors $\{b^\star_k\}$, and the minimum computation MSE of problem \eqref{pro6} are give as
\begin{alignat}{3}
& g^\star=\sqrt{\frac{1}{P}{\sum_{k=1}^{K} \left(\frac{P h_k}{\sigma^2+Ph_k^2}\right)^2}}\label{astar},\\ \label{b_k}
& b_k^\star=\frac{Ph_k}{\sigma^2+Ph_k^2}\sqrt{\frac{P}{\sum_{k=1}^{K} (\frac{Ph_k}{\sigma^2+Ph_k^2})^2}}, \forall k\in\mathcal{K},\\
& \mathsf{MSE^\star}=\sum_{k=1}^{K}\frac{\sigma^2}{\sigma ^2+Ph_k^2}\label{msestar}.
\end{alignat}
\end{theorem}

\begin{remark}\label{remark:1}
\normalfont
We see that the optimal Rx-scaling factor monotonically decreases with the increasing sum-power limit $P$; while the optimal Tx-scaling factors monotonically increase with $P$.
Interestingly, from \eqref{b_k}, it is clear that both the sensors with poor and good channel conditions should use less power than the ones with moderate channel conditions.
Also, we see that the minimum $\mathsf{MSE}$ monotonically decreases with the increasing SNR $P/\sigma^2$ and the channel-power gains.

\end{remark}

\section{Optimal Sum Power with Computation-MSE Constraint}\label{sec:P2}
In this section, we consider the optimal scaling-factor design problem to minimize the sum power of the AirComp system under the constraint of computation MSE. 
The problem is formulated as
\begin{subequations}\label{pro19}
	\begin{alignat}{2}
	\label{P2_a}
	&\!\min_{g, \{b_k\}}      &\quad& {\mathsf{PW} =\sum_{k=1}^{K}\vert b_k \vert^2}\\ \label{P2_b}
	&\text{subject to} &      & \mathsf{MSE} = \sum_{k=1}^{K} \vert g h_k b_k -1 \vert^2 + \sigma^2 \vert g \vert^2 \leq  \epsilon,
	\end{alignat}
\end{subequations}
where $\epsilon$ is the computation-MSE limit. To avoid trivial problems, it is assumed that $\epsilon<K$. Otherwise, the optimal solution of problem \eqref{pro19} is $b_k=0,\forall k\in \mathcal{K}$.

Problem~\eqref{pro19} is non-convex due to the non-convexity of \eqref{P2_b}.
However, if $g$ is fixed, it is convex. When $g$ is fixed, the Karush-Kuhn-Tucker (KKT) conditions~\cite{boyd2004convex}, which are necessary conditions of the optimal
solution of problem, are obtained as
\begin{numcases}{}
\!\frac{\partial\! \left(\!\eqref{P2_a}\!+\! \lambda_2 (\sum_{k=1}^{K}\!\vert g h_k b_k \!-\!1 \vert^2 \!+\! \sigma^2 \vert g \vert^2 \!-\! \epsilon)\!\!\right)}{\partial b_k}\!=\!0\label{secondlimit},\\
\label{P2_equal}
\sum_{k=1}^{K}\vert g h_k b_k -1 \vert^2 + \sigma^2 \vert g \vert^2 - \epsilon \leq 0,\\
\lambda_2(\sum_{k=1}^{K}\vert g h_k b_k -1 \vert^2 + \sigma^2 \vert g \vert^2 - \epsilon)=0,
\lambda_2 \geq 0,
\end{numcases}
where $\lambda_2$ is the KKT multiplier. It can be verified that $\lambda_2>0$ and the equality of \eqref{P2_equal} holds.
From \eqref{secondlimit}, we further  have
\begin{equation}\label{b2}
b_k=\frac{\lambda_2gh_k}{1+\lambda_2g^2h_k^2}, \forall k \in \mathcal{K}.
\end{equation}
Then, \eqref{pro19} is converted as
\begin{subequations}
	\label{pro19_trans}
	\begin{alignat}{2}
		\label{pro19_trans_target}
	&\!\min_{g, \lambda_2}      &\qquad& {\sum_{k=1}^{K} (\frac{\lambda_2gh_k}{1+\lambda_2g^2h_k^2})^2},  \\
	\label{pro19_trans_const}
	&\text{subject to} &      & \sum_{k=1}^{K}\left( \frac{1}{1+\lambda_2 g^2h^2_k}\right)^2 + \sigma^2 \vert g \vert^2  = \epsilon,\\
	& & & \lambda_2 \textgreater 0.
	\end{alignat}
\end{subequations}	
Note that problem \eqref{pro19_trans} is still non-convex.
We introduce a sequence of variables $\{\tau_k\}$, where
\begin{equation}\label{uptau}
\tau_k=\frac{1}{1+\lambda_2g^2h_k^2} \in (0,1), \forall k \in \mathcal{K}.
\end{equation}
Taking \eqref{uptau} into \eqref{pro19_trans_const}, we have
\begin{equation}\label{g2'}
g^2=\frac{\epsilon-\sum_{k=1}^{K}\tau_k^2}{\sigma^2},
\end{equation}
Taking \eqref{uptau} and \eqref{g2'} into \eqref{pro19_trans_target}, problem~\eqref{pro19_trans} is equivalent~to
\begin{subequations}
	\label{pro19_trans'}
	\begin{alignat}{2}
	\label{pro19_trans_target'}
	&\!\min_{\{\tau_k\}}      &\qquad& \sigma^2{\left(\sum_{k=1}^{K}\frac{(1-\tau_k)^2}{h_k^2}\right)} \big/ {\left(\epsilon-\sum_{k=1}^{K}\tau_k^2\right)},  \\
		\label{pro19_trans_const'}
	&\text{subject to} & & \sum_{k=1}^{K}\tau^2_k < \epsilon,	\\
	\label{pro19_trans_const''}
	& & & 0< \tau_k <1, \forall k\in \mathcal{K}.
	\end{alignat}
\end{subequations}
It can be proved that \eqref{pro19_trans_target'} is convex within the region defined by \eqref{pro19_trans_const'} and \eqref{pro19_trans_const''}.
In what follows, we will show that the extreme point of function \eqref{pro19_trans_target'} locates in the constraint region.
Letting the partial derivative of \eqref{pro19_trans_target'} in terms of $\tau_k$ equal to zero, we have
\begin{equation}\label{tau}
\tau_k = \frac{1}{1+M/c_k}, \forall k \in \mathcal{K},
\end{equation}
where 
$c_k\triangleq 1/h^2_k$
and
\begin{equation}\label{M}
M\triangleq\frac{\sum_{k=1}^{K}c_k(1-\tau_k)^2}{\epsilon-\sum_{k=1}^{K}\tau_k^2}.
\end{equation}
Taking \eqref{tau} into \eqref{M}, $M$ can be obtained by solving the equation of
\begin{equation}
M=\frac{\sum_{k=1}^{K}c_k(\frac{M}{c_k+M})^2}{\epsilon-\sum_{k=1}^{K}(\frac{c_k}{c_k+M})^2},
\end{equation}
which can be proved to have a unique solution in the region $(0,\infty)$. Thus, $\tau_k$ in \eqref{tau} satisfies the constraints \eqref{pro19_trans_const'} and \eqref{pro19_trans_const''}.
From \eqref{b2}, \eqref{uptau} and \eqref{g2'}, we have the following results.
\begin{theorem}\label{theorem:2}
\normalfont
The optimal Rx-scaling factor $g^\star$ and the optimal  Tx-scaling factors $\{b^\star_k\}$, and the minimum sum power of problem \eqref{pro19} are give as
	\begin{alignat}{2}\label{amsestar}
	& g^*=\frac{1}{\sigma}{\sqrt{\epsilon-\sum_{k=1}^{K}(\frac{1}{1+Mh_k^2})^2}},\\
	& b_k^*=\frac{\sigma M h_k}{(1+Mh_k^2){\sqrt{\epsilon-\sum_{k=1}^{K}(\frac{1}{1+Mh_k^2})^2}}}, \forall k \in \mathcal{K}\\
	&\mathsf{PW}^\star = \sigma^2\!\left(\!{\sum_{k=1}^{K}\frac{(1\!-\!1/(1+\frac{M }{c_k}))^2}{h_k^2}}\!\right)\!/\!\left(\!{\epsilon\!-\!\sum_{k=1}^{K}1/(1+\frac{M }{c_k})^2}\!\right).
	\end{alignat}
\end{theorem}
\begin{remark}
	\normalfont
From Theorem~\ref{theorem:2}, it can be observed that the optimal Tx-scaling factors and the minimum sum power increase with the increasing receiver's noise power $\sigma^2$, while the Rx-scaling factor decreases with $\sigma^2$.	
Unlike the optimal computation-MSE policy in Theorem~\ref{theorem:1}, the optimal sum-power policy in Theorem~\ref{theorem:2} is more complex and cannot provide more insights directly in terms of the computation-MSE limit $\epsilon$ and the channel coefficients $\{h_k\}$. We will numerically demonstrate these properties in the following.
\end{remark}

\section{Numerical Results}
In this section, we present the numerical results for the optimal computation-MSE policy and the optimal sum-power policy of the AirComp system based on Theorems~\ref{theorem:1} and \ref{theorem:2}, respectively.
Unless otherwise stated, the number of sensor is $K=10$, the sensors' sum-power limit is $P=10$, the AirComp computation-MSE limit is $\epsilon =5$, the receiver's noise power is $\sigma^2=1$.
Also, we assume that $h_1<\cdots<h_K$.

\begin{figure}[t]
	\centering
	\includegraphics[scale=0.6]{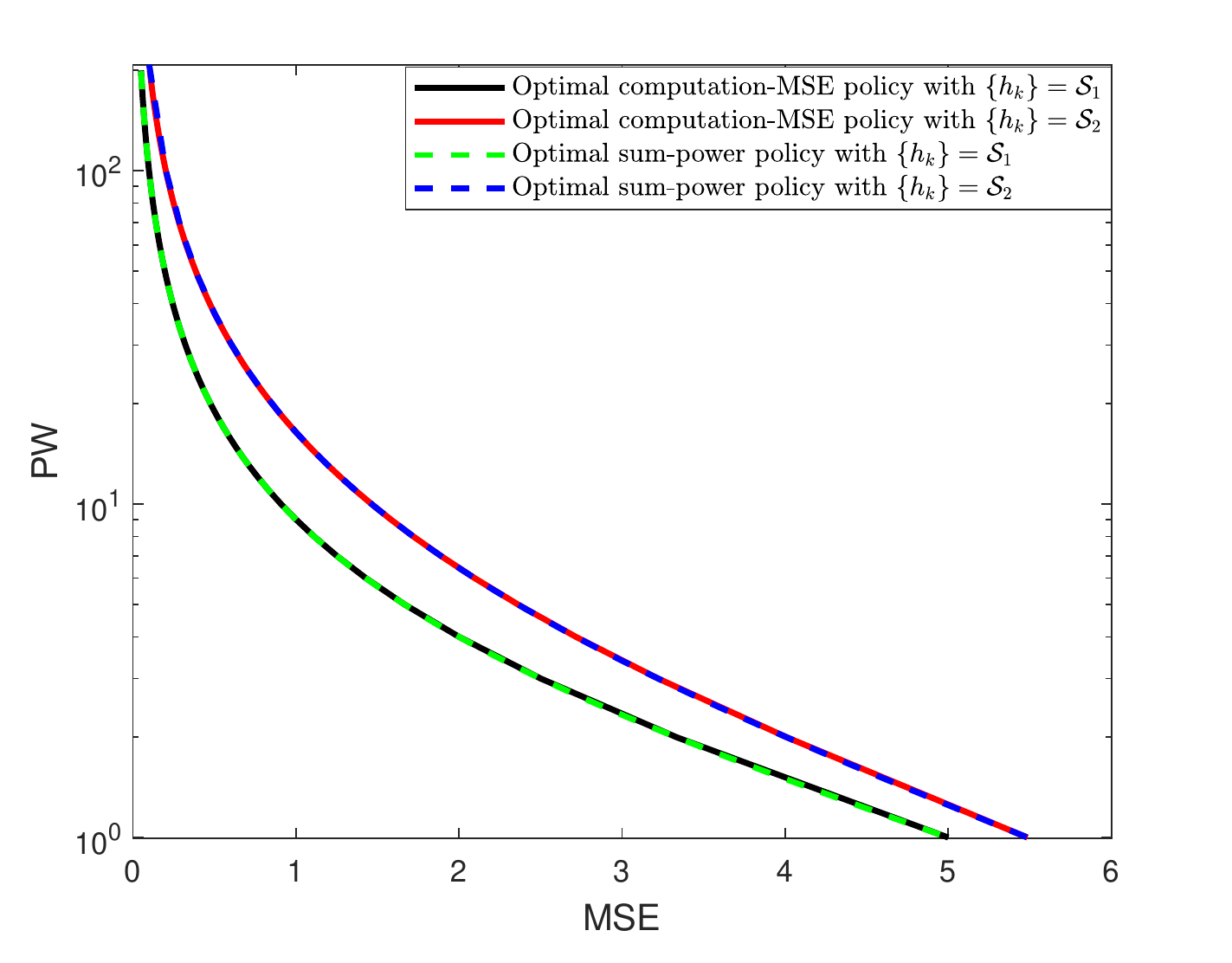}
	\vspace{-0.8cm}
	\caption{$\mathsf{PW}$ versus $\mathsf{MSE}$.}
	\label{fig:dual}
    \vspace{-0.5cm}
\end{figure}

\begin{figure}[t]
	\centering
	\vspace{-0.1cm}
	\includegraphics[scale=0.6]{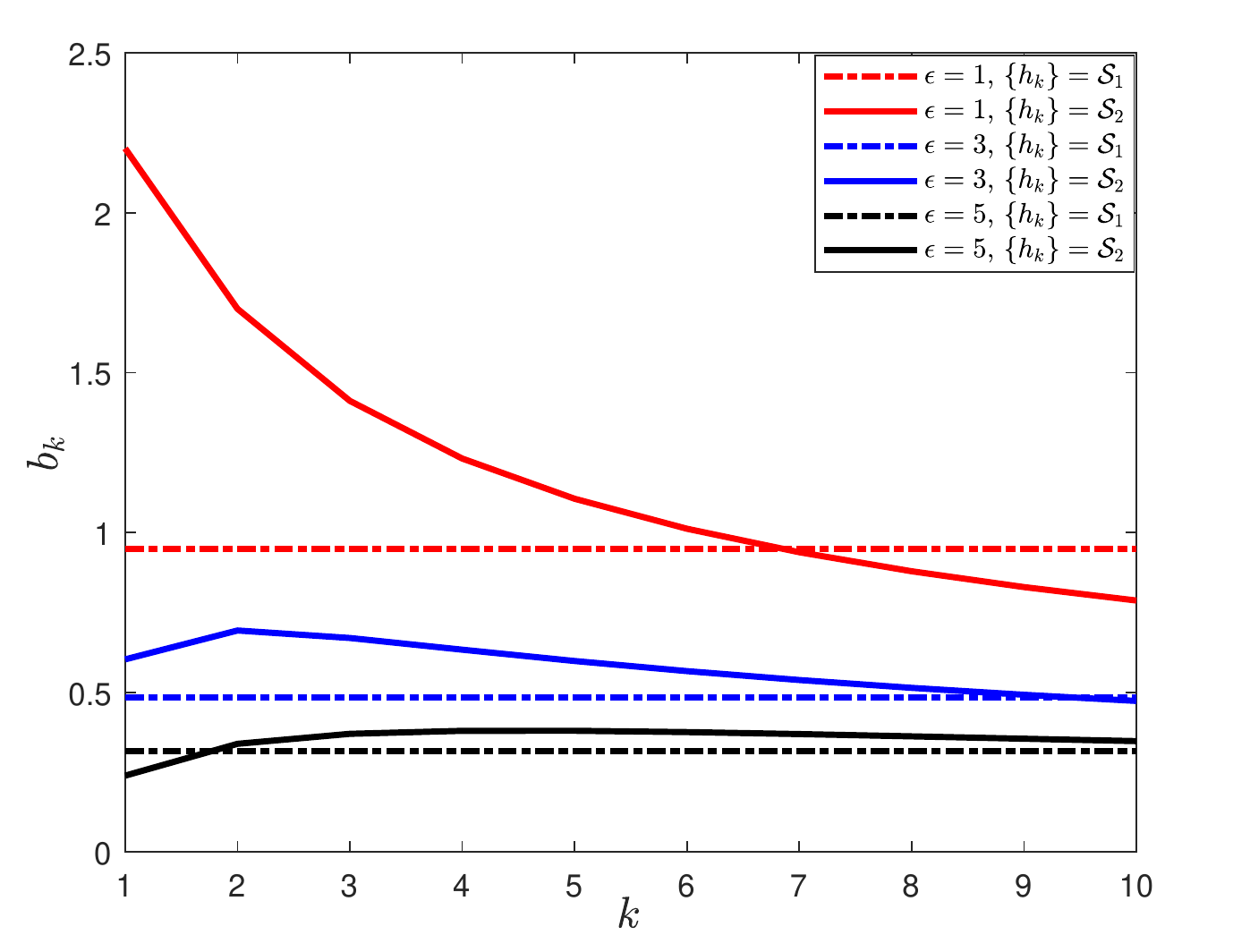}
		\vspace{-0.3cm}
	\caption{The Tx-scaling factors $\{b_k\}$ of the optimal computation-MSE policy.}
	\label{fig:bimse}
		\vspace{-0.5cm}
\end{figure}

\begin{figure}[t]
	\centering
	\includegraphics[scale=0.6]{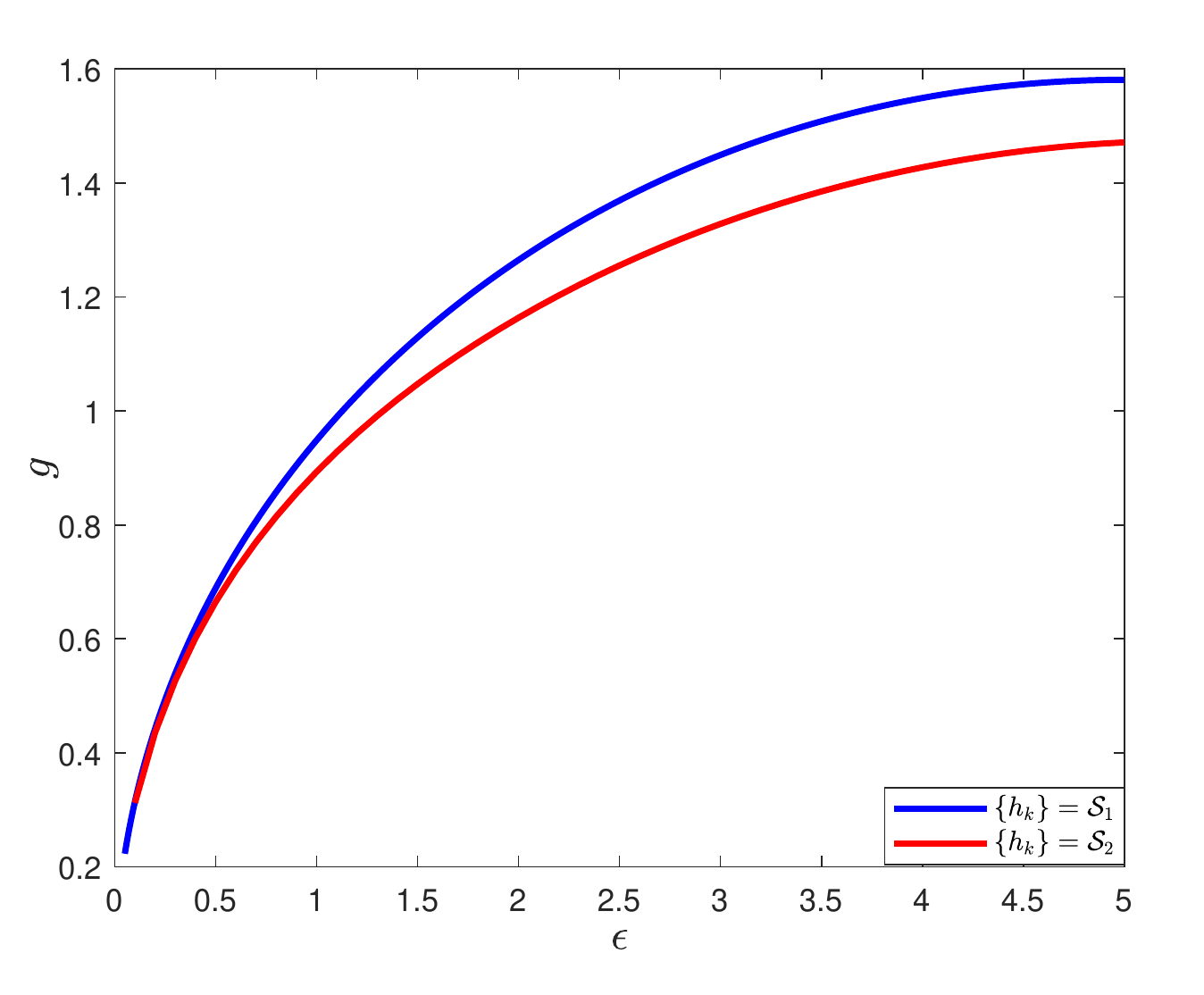}
	\vspace{-0.5cm}	
	\caption{The Rx-scaling factor $g$ versus computation-MSE constraint $\epsilon$.}
	\label{fig:amse}
	\vspace{-0.5cm}	
\end{figure}

In Fig.~\ref{fig:dual}, we plot the optimal MSEs under different constraints of the sum power and the optimal sum power under different MSE constraints, with different sets of channel-power gains, i.e., $\mathcal{S}_1=\{1,1,\cdots,1\}$ and $\mathcal{S}_2=\{0.1,0.3,0.5,0.7,0.9,1.1,1.3,1.5,1.7,1.9\}$.
We see that the relations between the optimal MSE versus the optimal sum power are the same in two different problems investigated in Sections~\ref{sec:P1} and~\ref{sec:P2} as expected, which also verifies the correctness of Theorems~\ref{theorem:1} and~\ref{theorem:2}.
Since the properties of the optimal computation-MSE policy have been directly obtained in Remark~\ref{remark:1}, we only present the numerical results for the optimal sum-power policy in Figs.~\ref{fig:bimse} and~\ref{fig:amse}.

In Fig.~\ref{fig:bimse}, we plot the optimal Tx-scaling factors $\{b_k\}$ of the optimal sum-power policy with different computation-MSE constraints $\epsilon$ and different channel coefficients.
It can be observed that a smaller $\epsilon$ leads to a larger sequence of Tx-scaling factors $\{b_k\}$.
We see that unlike the constant power allocation policy of the ideal case with identical channel coefficients (i.e., $\{\vert h_k \vert^2\}=\mathcal{S}_1$), the allocated power $|b_k|^2$ with non-identical channel coefficients (i.e., $\{\vert h_k \vert^2\}=\mathcal{S}_2$) first increases and then decreases with the channel-power gain $|h_k|^2$ when the computation-MSE constraint is loose, i.e., $\epsilon \geq 3$. Also, we see that the optimal power allocation policy approaches to a channel-inversion-like policy when the computation-MSE constraint is tight, i.e., allocating more power to the sensors with worse channel conditions.

In Fig.~\ref{fig:amse}, we plot the optimal Rx-scaling factors $g$ of the optimal sum-power policy versus the computation-MSE limit $\epsilon$ with  different channel coefficients.
We see that $g$ monotonically increases with the computation-MSE limit $\epsilon$. 
It can be observed that the Rx-scaling factor of the identical channel coefficient case is larger than that of the non-identical channel coefficient case.

We also investigate the performance of the optimal computation-MSE policy and the optimal sum-power policy of the AirComp system under independent and identically distributed (i.i.d.) Rayleigh fading channels, with different number of sensors $K$.
Intuitively, the computation MSE and the sum power of the AirComp system increase with the number of sensors. For fare performance comparison with different $K$, in the following, we present the results of normalized average MSE and sum power as $\mathsf{E}[\mathsf{MSE}]/K$ and $\mathsf{E}[\mathsf{PW}]/K$, respectively,
where the average is 
evaluated by Monte Carlo simulation with $10^6$ random channel realizations.
The sum-power limit and the computation-MSE limit are $P =10 K$ and $\epsilon = 0.2 K$, respectively.

\begin{figure}[t]
	\centering
	\includegraphics[scale=0.6]{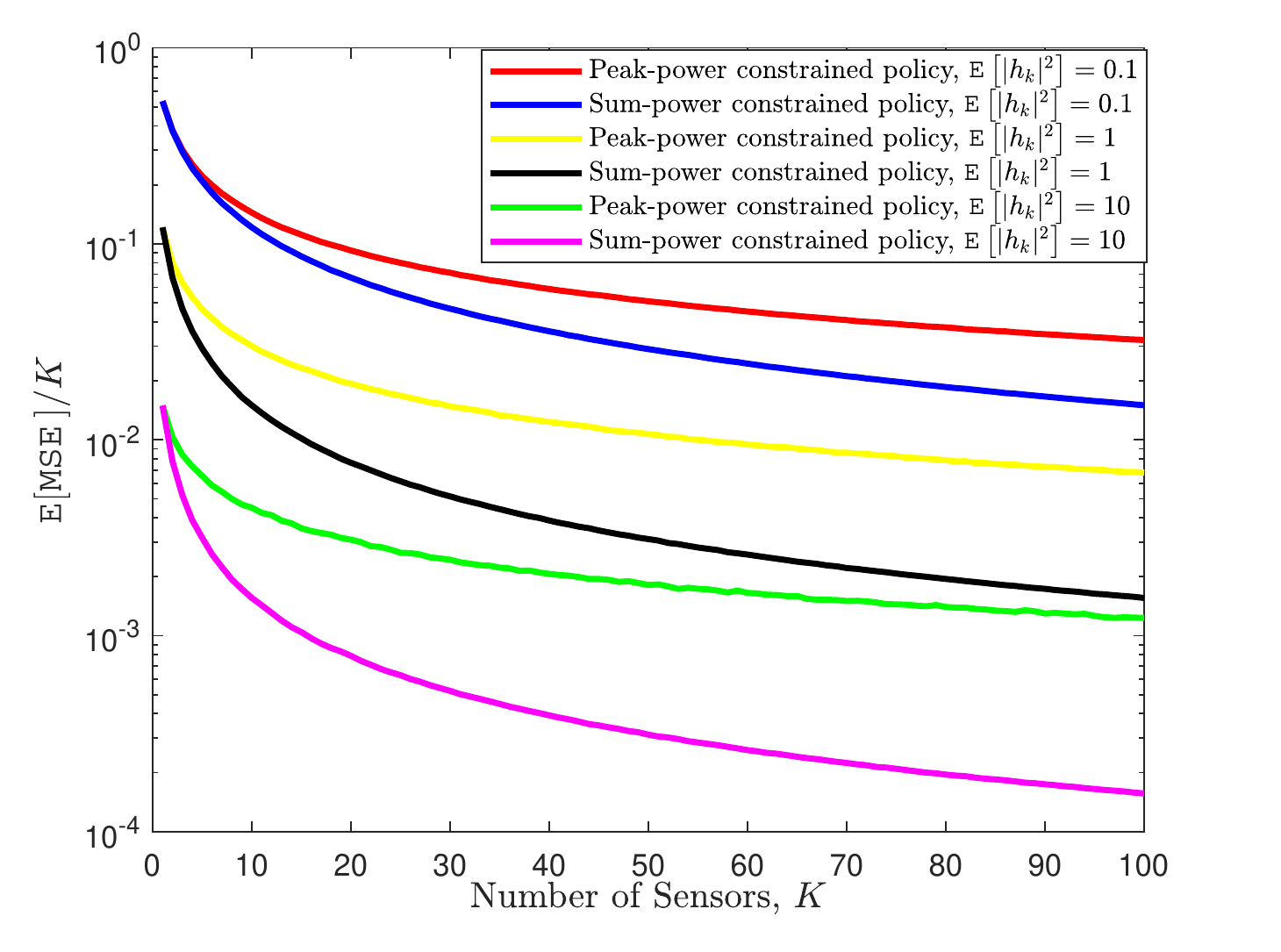}
	\vspace{-0.6cm}	
	\caption{The average computation MSE versus $K$.}
	\label{fig:msek}
	\vspace{-0.5cm}	
\end{figure}

In Fig.~\ref{fig:msek}, we plot the average MSE versus the number of sensors $K$ with different average channel-power gains and different AirComp policies, i.e., the optimal peak-power constrained policy~\cite{liu2019over}, where the peak power constraint is $10$, and the optimal sum-power constrained policy in Section~\ref{sec:P1}. It is clear that the average MSE decreases with the increasing $K$ and the average channel power gain $\mathsf{E}[|h_k|^2]$ in both the policies.
Also, we see that the optimal policy under the sum-power constraint leads to a significantly smaller computation MSE than that of the peak-power constraint optimal policy and the gap increases with $K$, due to the additional flexibility in power allocation.

We have also plotted figures about the average sum power versus the number of sensors $K$ with different average channel-power gains of the optimal sum-power policy in Section~\ref{sec:P2} (figures are not shown in the paper due to the space limitation). It can be observed that the average sum power decreases with the increasing $K$ and the increasing average channel power gain.

\section{Conclusions}
In the paper, we have proposed and solved the optimal computation-MSE problem and also the optimal sum-power problem of the AirComp systems, and have obtained closed-form solutions.
Our results have shown that for both policies, the sensors with poor and good channel conditions should use less power than the ones with moderate channel conditions. 



\end{document}